\documentclass[showpacs,twocolumn,prl,aps]{revtex4}
\usepackage{graphicx}

\begin{document}

\title[Short title for running header]{Absence of topological degeneracy in the Hubbard model on honeycomb lattice}
\author{Tao Li}
\affiliation{ Department of Physics, Renmin University of China,
Beijing 100872, P.R.China}
\date{\today}

\begin{abstract}
It is shown that the unique sign structure of the ground state of
the Hubbard model on honeycomb lattice, which is shown to be
insensitive to the trapped $Z_{2}$ gauge flux when the system is
defined on a torus, may cause the absence of topological degeneracy
on this bipartite system. Examples of variational Mott insulating
state on the honeycomb lattice are given to illustrate the close
relation between the sign structure of the ground state and the
(absence of) topological degeneracy.
\end{abstract}
\maketitle

The search for spin liquid ground state is a central issue in the
study of the strongly correlated electron systems. The spin liquid
state represents a novel state of matter beyond the Landau-Ginzburg
description and supports new kinds of order and excitation. It is
generally believed that the study of spin liquid state will not only
deepen our understanding on the organizing principles of condensed
matter systems, but also result in new applications that is not
possible from conventional materials. In particular, the topological
order and the related fractionalized excitations are proposed to
implement the key steps of topological quantum computation.

It is generally believed that the geometrically frustrated quantum
antiferromagnetic systems are ideal places to find spin liquid
ground state. For this reason, the study of Heisenberg model on
Kagome lattice and triangular lattice have received considerable
interests. More recently, the spin liquid states are also proposed
to appear in Hubbard models on frustrated lattices as a result of
the multi-spin exchange processes. The bipartite lattice, on which
the antiferromagetic exchange interaction is not frustrated, is
generally not believed to be favorable for the formation of the spin
liquid ground state. For example, on the square lattice, the system
develops magnetic long range order as soon as one turns on the
electron correlation. With these understandings in mind, it is quite
unexpected that a spin liquid ground state can exist in the Hubbard
model on the bipartite honeycomb lattice.

According to the numerical results reported recently by Meng et.
al\cite{Meng}, a state with full gaps to both charge and spin
excitations and full symmetry of the Hamiltonian emerges in a small
interaction range of $3.5<U/t<4.2$ for the Hubbard model on
honeycomb lattice. This state intervenes between a semimetal phase
with a Dirac-type spectrum for $U/t<3.5$ and an antiferromagnetic
ordered phase with spin wave excitation for $U/t>4.2$, resulting in
a counterintuitive non-monotonic evolution of the spin excitation
spectrum with $U/t$. What makes this state even more special is that
although it has a full gap to spin excitation, it has no
accompanying topological degeneracy.

In a loose sense, a spin liquid state can be defined as a quantum
disordered insulting state of a many electron system that respect
all the symmetries of the Hamiltonian. However, to exclude the
trivial case of a band insulator, in which the spin degree of
freedoms cancel out in each unit cell, one should add the further
requirement that each unit cell of the lattice contains an odd
number of electrons. A system with an odd number of electrons in
each unit cell will inevitably posses gapless excitations at the
mean field level if there is no symmetry breaking mechanism to
enlarge the unit cell\cite{Oshikawa1}. Built on the higher
dimensional generalization of the celebrated Lieb-Schultz-Mattis
theorem in one dimension, it is now generally believed that a spin
liquid state should either be gapless, or, while possessing a bulk
gap to spin excitation, show topological degeneracy\cite{Hastings}.
The topological degeneracy denotes the degeneracy between ground
states that can not be distinguished locally but are globally
distinct. The existence of topological degeneracy is one of the most
important manifestation of topological order, which is believed to
be a prerequisite for the emergence of fractionalized excitations in
the spin liquid background\cite{Oshikawa2}. For these reasons, it is
not surprising that the negative result on the topological
degeneracy reported by Meng et.al. in the spin liquid state has
aroused much interests and also confusions in the
community\cite{Sun,Xu,Wang,Lu}.

In a sense, the absence of topological degeneracy in the Hubbard
model on honeycomb lattice should not be that surprising since the
honeycomb lattice is a complex lattice with two inequivalent sites
in each unit cell. A state with one electron per site on average
actually corresponds to integer filling rather than half filling.
Thus the above mentioned hypothesis on the topological degeneracy
should not apply here. However, it is not clear to what extent is
the oddness of electron number in each unit cell essential for the
existence of the topological degeneracy in a fully gaped system. As
the topological degeneracy is the key to the exoticness of a spin
liquid\cite{Oshikawa2} and as some doubts have been raised on the
negative results of topological degeneracy\cite{Wang,Lu} in the
numerical work, it is valuable to understand if Hubbard model on
honeycomb can really support topological degeneracy. Even leaving
apart the topological degeneracy, it is still a challenging task to
compose a picture for the spin liquid state reported in the
numerical works and to understand how a full gap can be generated in
the spin excitation spectrum without breaking any
symmetry\cite{Sun,Xu,Wang,Lu}.

In this paper, we provide an analytical argument for the absence of
topological degeneracy in the Hubbard model on honeycomb lattice.
This argument is based on the unique sign structure of the ground
state of this bipartite system, which is shown to be insensitive to
the trapped gauge flux when the system is defined on a torus. To
illustrate the effect of the ground state sign structure on the
topological degeneracy, we also present the results of a variational
study on some Mott insulating states on the honeycomb lattice.

The model studied in this paper reads
\begin{equation}
\mathrm{H}=-t\sum_{<i,j>,\sigma}(c^{\dagger}_{i,\sigma}c_{j,\sigma}+h.c.)+U\sum_{i}n_{i\uparrow}n_{i\downarrow},
\end{equation}
in which the first sum is over nearest neighboring sites on the
honeycomb lattice. The honeycomb lattice is a complex lattice with
two inequivalent sites in each unit cell(see Fig.1) and can be
divided into two sublattices(even and odd). The hopping terms are
nonzero only between sites on different sublattices. With
$\vec{b}_{1}$ and $\vec{b}_{2}$ denoting the primitive lattice
vectors in reciprocal
space($\vec{b}_{i}\cdot\vec{a}_{j}=2\pi\delta_{i,j}$), the momentum
in the Brillouin zone can be parameterized as
$\vec{k}=k_{1}\vec{b}_{1}+k_{2}\vec{b}_{2}$. The kinetic part of the
Hamiltonian can then be diagonalized as follows
\begin{equation}
\mathrm{H}_{0}= \sum_{k\sigma,s}E_{k,s}\gamma^{ \dagger}_{
k\sigma,s}\gamma_{k\sigma,s},
\end{equation}
in which $E_{k,s}=s|1+e^{ik_{1}}+e^{ik_{2}}|$ and $s=\pm 1$.
$\gamma_{k\sigma,s}$ is corresponding eigen-operator. The bare
dispersion has two Dirac points at
$(k_{1},k_{2})=\pm(2\pi/3,-2\pi/3)$ in the first Brillouin zone.
When the electron density is such that there is one electron per
site on average, the Fermi surface of the free system shrinks into
these two Dirac points. Thus, although the honeycomb lattice is
bipartite and the antiferromagnetic interaction is not frustrated, a
finite strength of local correlation is needed to induced the
instability of spin density wave ordering.
\begin{figure}[h!]
\includegraphics[width=8cm,angle=0]{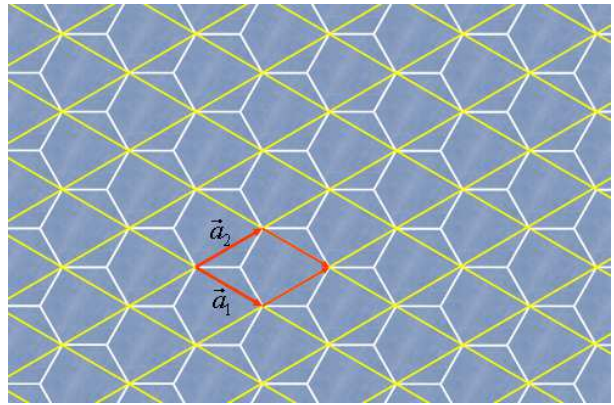}
\caption{The honeycomb lattice and its unit cell.} \label{fig1}
\end{figure}

It should be noted that the gaplessness of the bare dispersion at
$(k_{1},k_{2})=\pm(2\pi/3,-2\pi/3)$ is protected by the bipartite
nature and the three-fold rotational symmetry of the honeycomb
lattice. Supposing that the hopping integral is nonzero only between
sites on different sublattices, then by the three-fold rotational
symmetry, a hopping term between sites separated by an arbitrary
vector
$\vec{R}_{1}=(m+\frac{1}{3})\vec{a}_{1}+(n+\frac{1}{3})\vec{a}_{2}$
is always accompanied by two other hoppings of the same amplitudes
at separations
$\vec{R}_{2}=-(m+n+\frac{2}{3})\vec{a}_{1}+(m+\frac{1}{3}\vec{a}_{2})$
and
$\vec{R}_{3}=(n+\frac{1}{3})\vec{a}_{1}-(m+n+\frac{2}{3})\vec{a}_{2}$.
It is then easy to check that the expression $\sum_{i=1,2,
3}e^{i\vec{k}\cdot\vec{R}_{i}}$ is identically zero at
$(k_{1},k_{2})=\pm(2\pi/3,-2\pi/3)$. The same arguments can also be
used to show that BCS pairing between sites on the different
sublattices and respecting the three-fold rotational symmetry can
neither open a gap at $(k_{1},k_{2})=\pm(2\pi/3,-2\pi/3)$.

With this in mind, it is then quite unusual that a full gap in both
spin and charge sector can be opened without any symmetry breaking.
Could it be possible that the system spontaneously violates the
bipartite nature of the honeycomb lattice by generating hopping or
pairing terms between sites on the same sublattice\cite{Wang,Lu}?
The analytical argument present below suggests that this is very
unlikely to be the case. The same argument also provides an
understanding on the absence of the topological degeneracy in the
Hubbard model on honeycomb, or more generally, bipartite lattice.

Our argument is based on the now well known Lieb's theorem on the
sign structure of the ground state on bipartite lattice for the
Hubbard model. On a bipartite lattice such as the honeycomb lattice
studied in this paper, the Hubbard model is particle-hole symmetric.
To be more specific, through the following unitary transformation
\begin{equation}
\left( {\begin{array}{c}
c_{i\uparrow}\\
c_{i\downarrow}\\
\end{array}}
\right) \longrightarrow \left( {\begin{array}{c}
c_{i\uparrow}\\
\eta_{i}c^{\dagger}_{i\downarrow}\\
\end{array}}
\right),
\end{equation}
in which $\eta_{i}=-1$ for $i$ in the odd sublattice and is
otherwise 1, the Hubbard model takes the form
\begin{equation}
\mathrm{H}=-t\sum_{<i,j>,\sigma}(c^{\dagger}_{i,\sigma}c_{j,\sigma}+h.c.)-U\sum_{i}n_{i\uparrow}n_{i\downarrow},
\end{equation}
up to a chemical potential term. Under such a transformation, a
repulsive Hubbard model is mapped into an attractive Hubbard model.
At the same time, the spin and charge degree of freedoms interchange
their roles in the particle-hole transformed system. For example, a
repulsive system with one electron per site is mapped into an
attractive system with zero magnetization.

In a celebrated paper appeared in 1989 by Lieb\cite{Lieb}, it is
proved that the attractive Hubbard model Eq.(4) on any lattice has a
unique ground state with a well defined sign structure. More
specifically, if we use the Fock basis
$|\alpha,\uparrow\rangle=\prod_{k}c^{\dagger}_{i_{\alpha
k}\uparrow}|0\rangle$ and
$|\alpha,\downarrow\rangle=\prod_{k}c^{\dagger}_{i_{\alpha
k}\downarrow}|0\rangle$ to expand the ground state as
$|\Psi\rangle=\sum_{\alpha,\beta}\phi_{\alpha,\beta}|\alpha,\uparrow\rangle\otimes|\beta,\downarrow\rangle$,
then the coefficient matrix $\phi$(whose matrix elements are
amplitudes of the ground state in the Fock basis) can be shown to be
positive definite, although in general its matrix elements can be
negative as a result of the Fermion sign. This can be understood as
a result of the local attractive interaction which encourages the
down spin electrons to follow the trail of the up spin electrons. In
particular, the ground state amplitude is positive definite when
electrons of both spins share the same set of lattice sites. In term
of the repulsive Hubbard model before the particle-hole
transformation, this corresponds to a singly occupied configuration.
Keep in mind the alternative sign $\eta_{i}$ in the particle-hole
transformation Eq.(3), it can be shown easily that in the repulsive
Hubbard model, the ground state amplitudes in the singly occupied
subspace satisfy the so called Marshall sign rule. The Marshall sign
rule claims that the ground state amplitudes in the singly occupied
subspace are real and their signs are given by
$(-1)^{N^{even}_{\downarrow}}$ up to a global phase factor. Here
$N^{even}_{\downarrow}$ is the number of down spin electrons in the
even sublattice.

The existence of such a sign structure in the ground state, which is
a result of bipartite nature of the honeycomb lattice, makes it very
unlikely that the reported spin gap is generated by intra-sublattice
hopping or pairing terms\cite{Wang,Lu}. At the same time, the sign
structure of the ground state provides a way to understand the
absence of the topological degeneracy in this system. To appreciate
this point, it is crucial to realize that Lieb's proof is general
enough to address not only the sign structure of the ground state,
but also it response to the trapped gauge flux when the system is
defined on a torus. Lieb's proof applies whenever the lattice is
bipartite, while up to a gauge transformation trapping a gauge flux
in the holes of the torus amounts simply to change the phase of the
hopping integrals on the bonds across the boundary of the system.
Thus, the ground state after the gauge flux trapping has exactly the
sign structure as that before the gauge flux trapping up to an
inessential global phase. It is just this boundary condition
independence of the sign structure of the ground state that enables
us to arrive at our conclusion.

To be more specific, the discussion below will be restricted to the
topological degeneracy of the $Z_{2}$ type, which is the most
commonly envisaged in a gapped quantum spin liquid\cite{Wen}. A
scheme to detect topological degeneracy in the resonating valance
bond(RVB) state generated from Gutzwiller projection of BCS mean
field state is proposed by Ivanov and Senthil in 2002\cite{Ivanov}.
They proposed to check the orthogonality between RVB states with
different number of trapped visons(the $Z_{2}$ topological
excitation)in the holes of a multiply connected manifold. For a
topological ordered system, states with different number of trapped
visons in the holes are orthogonal to each other and form
topological degeneracy. On the other hand, for a topological trivial
state, the visons can tunnel through the bulk of the system and
escape from the holes. The topological degeneracy is lifted by such
tunneling events. Built on the conceptual link in the RVB picture
between a quantum spin liquid and a quantum phase disordered
superconductor, Ivanov and Senthil also proposed to construct the
vison excitation from quantized superconducting vortex in the mean
field state. It is then found that as a result of the $Z_{2}$
character of the quantized superconducting vortex, trapping a vison
in a hole of multiply connected manifold for the RVB state amounts
to changing the boundary condition around the hole from periodic to
anti-periodic or vice versa.

In a later generalization by the present author and Yang\cite{Li},
the role of the superconducting vortex in defining the vison is
replaced by that of a $Z_{2}$ gauge flux. Thus, to detect
topological degeneracy in the RVB state, one should check the
sensitivity of the ground state to a quantized gauge flux in the
holes on a multiply connected manifold. Such a definition closely
resembles the way that Kohn chose to define a insulator\cite{Kohn}.
According to this definition, the key difference between an
insulator showing topological degeneracy(order) and a trivial
insulator is that the former can respond nonlocally to a trapped
$Z_{2}$ gauge flux while the latter has no such nonlocal response.

In \cite{Li}, it is also shown that RVB states generated from
bipartite mean field Hamiltonian through Gutzwiller projection all
satisfy the Marshall sign rule, no matter what is the gauge
structure of the corresponding effective theories at the Gaussian
level. As a result of such sign structure, it is proved generally
that all RVB state generated from a bipartite mean field Hamiltonian
can not support topological order. The same argument can also be
adopted here after some modifications. The modification is necessary
for two reasons. First, here we are discussing ground state of a
given Hamiltonian rather than a general RVB state with unspecified
Hamiltonian. Second, the model we discuss here also has charge
excitation, the existence of which will inevitably cause Fermion
sign and the ground state wave function has a definite sign only in
the subspace of singly occupied state.

To deal with the first problem, we simply introduce the $Z_{2}$
gauge flux directly in the Hamiltonian rather than in the effective
theory as was done in the previous study. From the above argument on
the topological degeneracy, this seems to be intuitively even more
appealing as it allow us to probe directly the response of the
ground state to the trapped $Z_{2}$ gauge flux.

Now we check the orthogonality of the ground states with and without
a trapped $Z_{2}$ gauge flux in the holes of a multiply connected
manifold. Let $\phi$ and $\phi^{'}$ be the wave function matrixes of
the ground states $|\Psi\rangle$ and $|\Psi^{'}\rangle$ before and
after the flux trapping, then as the sign structure of the ground
state is insensitive to the trapped gauge flux, both matrixes should
be positive definite up to a global phase factor. The overlap of the
two states is given by
$\langle\Psi|\Psi^{'}\rangle=\mathrm{Tr}\phi\phi^{'}$, in which we
have used the fact that both $\phi$ and $\phi^{'}$ can be assumed to
be Hermitian without loss of generality. In the diagonal
representation of $\phi$, the overlap can be written as
$\langle\Psi|\Psi^{'}\rangle=\sum_{i}w_{i}\phi^{'}_{i,i}$, in which
$w_{i}$ are the positive definite eigenvalues of the matrix $\phi$
and $\phi^{'}_{i,i}$ are the diagonal matrix elements of matrix
$\phi^{'}$ is this representation. Since $\phi^{'}$ is also positive
definite, $\phi^{'}_{i,i}>0$. Thus the overlap is the sum of
positive terms and is nonzero for finite system.

To prove the absence of the topological degeneracy, one should still
go to the thermodynamic limit to see if the overlap converges to a
finite value. At present a proof of this is beyond our reach. With
this in mind, the results presented above can be either interpreted
as a suggestive evidence for the absence of the topological
degeneracy, or, on the other way around, an excuse for the failure
to detect the anticipated topological degeneracy on finite system.
However, forbiddenness of topological degeneracy induced by the
Marshall-sign structure of ground state has been illustrated before
at the variational level for various RVB states and the same thing
can also be done for the honeycomb system.

For this purpose, we investigate two typical Mott insulating states
on the honeycomb lattice with built-in Marshall sign structure. Both
of these states are generated by Gutzwiller projection of BCS mean
field states. The first is the Gutzwiller projection of the free
Fermion state on the honeycomb lattice with nearest neighboring
hopping terms and two Dirac points at
$(k_{1},k_{2})=\pm(2\pi/3,-2\pi/3)$. The second is the Gutzwiller
projection of the BCS pairing state with $d+id'$ pairing pattern
between neighboring sites\cite{Did}. The $Z_{2}$ gauge flux can be
imposed in the boundary condition of the BCS mean field
Hamiltonian\cite{Ivanov,Li}.

As a result of a general theorem proved in \cite{Li}, both states
satisfy the Marshall sign rule, even if the unprojected wave
function is not real(as for the $d+id'$ state). To see if
topological degeneracy can survive in these states, we calculate the
overlap of the states defined on a torus with different number of
$Z_{2}$ gauge flux threaded in both holes of the torus. The overlap
between the wave functions can be easily calculated by with the
variational Monte Carlo method. The results for the projected free
Fermion state is presented in Fig. 2, in which we have shown the
overlap between the state with no flux in both holes and the state
with $Z_{2}$ flux in one hole and no flux in the other hole of the
torus. The calculation is done on lattice with $L\times L \times 2$
sites and $L$ is so chosen to avoid the Dirac nodes in the momentum
space. The results rapidly converges to a finite value when $L>10$ .

\begin{figure}[h!]
\includegraphics[width=8cm,angle=0]{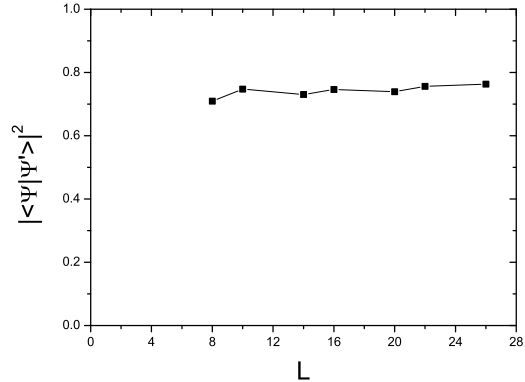}
\caption{The overlap between states with different number of $Z_{2}$
gauge flux in both holes of the torus for the Gutzwiller projected
free Fermion state on honeycomb lattice. $|\Psi\rangle$ denotes the
state with no gauge flux in both holes of the torus and
$|\Psi'\rangle$ denotes the state with $Z_{2}$ gauge flux in one
hole but no flux in the other hole of the torus.} \label{fig2}
\end{figure}

An even more striking example of how the sign structure of the
ground state can cause the absence of the topological degeneracy is
provided by the projected $d+id'$ wave pairing state on the
honeycomb lattice. Here, the unprojected mean field state break the
time reversal symmetry and has a complex wave function. In terms of
the effective field theories\cite{Wen}, the RVB state generated from
such a mean field ansatz would posses a $Z_{2}$ gauge structure and
thus support topological order. However, after the Gutzwiller
projection , the wave function becomes real up to a global phase and
satisfy the Marshall sign rule. In Fig.3 we present the overlap of
the state with $Z_{2}$ gauge flux in the hole surround by the
x-circumference but no flux in the hole surrounded by the
y-circumference(denoted as $|\Psi_{10}\rangle$) and the state with
$Z_{2}$ gauge flux in the hole surround by the y-circumference but
no flux in the hole surrounded by the x-circumference(denoted as
$|\Psi_{01}\rangle$). The overlap is also seen to converge to a
finite value in the thermodynamic limit.

\begin{figure}[h!]
\includegraphics[width=8cm,angle=0]{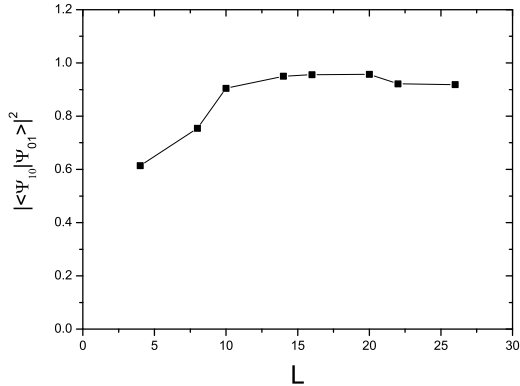}
\caption{The overlap between states with different number of $Z_{2}$
gauge flux in both holes of the torus for the Gutzwiller projected
$d+id'$ pairing state on honeycomb lattice. $|\Psi_{10}\rangle$
denotes the state $Z_{2}$ gauge flux in the hole surround by the
x-circumference but no flux in the hole surrounded by the
y-circumference and $|\Psi_{01}\rangle$ denotes the state with
$Z_{2}$ gauge flux in the hole surround by the y-circumference but
no flux in the hole surrounded by the x-circumference.} \label{fig3}
\end{figure}

Although both variational states presented above are not meant to
describe the spin liquid state found in the numerical simulation of
\cite{Meng}, they do illustrate well how the sign structure of the
ground state can cause the non-orthogonality of the states with and
without trapped gauge flux in the thermodynamic limit and thus the
forbiddenness of the topological degeneracy of the system. A full
description of the spin liquid state found in \cite{Meng} is beyond
the scope of this paper and requires obviously further refinement of
the variational wave function. For example, the virtual charge
fluctuation should obviously be taken into account. However, we
think these modifications will not change our conclusion
qualitatively.

In summary, we have presented both analytical arguments and
variational examples that suggest the topological degeneracy can not
survive in the Hubbard model on honeycomb lattice as a result of the
unique sign structure of its ground state. The same sign structure
is believed to be important also for other aspects of the ground
state and should be taken into account in the variational study.

The author is grateful to Fan Yang and Hong-Yu Yang for drawing his
attention to the work reported in \cite{Meng}. He also acknowledges
the valuable discussions with Ning-Hua Tong. This work is supported
by NSFC Grant No. 10774187 and National Basic Research Program of
China No. 2007CB925001.


\begin{thebibliography}{9}
\bibitem {Did}In the $d+id'$ pairing state, the pairing potentials on the bonds connecting the three nearest neighboring sites with a given site have the same amplitude, but differ in phase by $\pm\frac{2\pi}{3}$ with each
other. In our calculation, the ratio between the amplitude of the
pairing potential and the hopping integral $\frac{\Delta}{t}$ is set
to 0.1.
\end{thebibliography}
\end{document}